\documentclass{optica-article}

\journal{opticajournal} 

\articletype{Research Article}

\begin{document}

\title{A Novel Noise Analysis Method for Frequency Transfer System by Using ADEV Combine with EMD-WT}

\author{Xuan Yang,\authormark{1} Junhui Li,\authormark{1} Bin Luo,\authormark{1, *} Ziyang Chen,\authormark{2} and Hong Guo\authormark{2}}

\address{\authormark{1}State Key Laboratory of Information Photonics and Optical Communications, Beijing University of Posts and Telecommunications, Beijing 100876, China\\
\authormark{2}State Key Laboratory of Advanced Optical Communication Systems and Networks, School of Electronics, and Center for Quantum Information Technology, Peking University, Beijing 100871, China\\
}

\email{\authormark{*}luobin@bupt.edu.cn} 


\begin{abstract*} 
In precision frequency transfer systems, stringent requirements are imposed on the phase stability of transmitted signals. Throughout the transmission process, the inherent challenges of long-haul signal propagation inevitably introduce multiple noise components, including but not limited to thermal noise, phase fluctuations, and environmental interference. The system incline to use the conventional evaluation index – Allan deviation (ADEV) to reflect the system stability in order to evaluate the noise level. Whereas, ADEV can only provide numerical expression and lacks the time-frequency details. Therefore, a complete evaluation system is required by the system. In this paper, we present a groundbreaking integration of ADEV and wavelet transformed empirical mode decomposition (EMD-WT), establishing a novel analytical framework that enables simultaneous characterization of noise types and time-frequency domain properties. This synergistic approach achieves unprecedented dual-domain resolution in noise discrimination in frequency transfer systems. 

\end{abstract*}

\section{Introduction}

 In frequency transfer systems, the evaluation of frequency instability is conventionally performed using the ADEV, a metric capable of characterizing noise types and their periodicities within the system [1]. However, ADEV exhibits inherent limitations in precisely identifying the spectral characteristics and temporal localization of noise components. While commonly adopted as a standard reference metric, its utility in providing comprehensive analysis of the system's noise characteristics remains constrained.  Consequently, exclusive reliance on ADEV analysis proves inadequate, necessitating supplementary investigation through time-frequency analysis (TFA) methodologies.
 
Given the highly nonlinear and nonstationary nature of optical signals in frequency transfer systems, the selection of appropriate TFA techniques becomes imperative for effective signal characterization. Within the domain of time-frequency analysis, the Hilbert-Huang transform (HHT) has gained substantial recognition as a robust method for processing nonlinear and nonstationary signals. This intrinsic compatibility with complex signal behaviors renders HHT particularly suitable for analyzing optical signal manifestations in frequency transfer systems.

Over the past two decades, the HHT has emerged as a prominent methodology in nonlinear and nonstationary signal processing. Initially proposed for seismic signal analysis, its efficacy in extracting time-frequency features from complex signals has been empirically validated [2]. Subsequent applications have extended its utility to meteorological signal characterization [3], and more recently, to bearing fault detection and diagnostics in mechanical systems [4]. The foundation of HHT lies in empirical mode decomposition (EMD), an adaptive algorithm that recursively decomposes signals into intrinsic mode functions (IMFs) ordered by descending frequency components. However, practical implementation in frequency transfer systems reveals critical limitations: mechanical vibrations and fiber hopping phenomena induce high-frequency noise contamination, resulting in signal discontinuities that provoke mode mixing artifacts. This phenomenon fundamentally compromises the accuracy of subsequent Hilbert spectral analysis [5].

The inherent limitations of mode mixing constrain the standalone application of HHT for precision frequency signal analysis. While HHT demonstrates unique capabilities for nonlinear/nonstationary signal processing, alternative methodologies such as wavelet transform (WT) offer complementary advantages for TFA [6,7]. Unlike HHT's data-driven EMD approach, WT avoids mode mixing through predefined wavelet basis functions. From a theoretical perspective, ADEV exhibits mathematical equivalence to the variance representation in wavelet domains under specific conditions [8]. Notably, ADEV's variance metric demonstrates enhanced resolution with reduced sampling intervals compared to conventional WT implementations [9]. Nevertheless, WT's fixed-scale decomposition architecture imposes inherent constraints: predetermined scale and translation coefficients limit adaptability to dynamic signal characteristics, potentially obscuring low-intensity noise components beneath spectral features [10].

Comparative analysis reveals that neither HHT nor WT singularly satisfies the rigorous demands of frequency transfer signal characterization. A hybrid EMD-WT framework emerges as a promising solution, synergistically combining HHT's adaptive decomposition with WT's mathematical rigor. However, significant methodological challenges persist due to fundamental paradigm differences: EMD operates as a posteriori adaptive decomposition, whereas WT employs a priori basis selection. This method necessitates systematic validation of algorithmic compatibility, a critical focus of subsequent investigation in this work. Precedent exists in rotating machinery vibration analysis, where EMD-based signal stratification coupled with wavelet threshold denoising has demonstrated enhanced noise suppression capabilities [11]. Through strategic mitigation of inherent limitations in both methodologies, the EMD-WT synthesis proves particularly suited for frequency transfer signal analysis.

The novel contribution of this work resides in the integration of ADEV metrics with the EMD-WT framework to establish a comprehensive noise characterization paradigm in frequency systems. While ADEV – essentially a second-order statistical structure function – remains confined to second-order noise analysis, its concurrent implementation with wavelet time-frequency representations enables multidimensional diagnostics. By generating ADEV profiles and wavelet spectrograms from multiple IMFs, comparative analysis of their complementary features (ADEV slope-based noise classification vs. wavelet-based spectral-temporal localization) facilitates component-level noise attribution within the frequency transfer system. This dual-modality approach significantly enhances diagnostic resolution compared to conventional single-method implementations.

\section{Method}

\subsection{Method of EMD-WT}

Nonlinear and nonstationary characteristics predominantly govern most signals observed in system outputs. In the past decade, WT has achieved widespread adoption in nonstationary signal processing, particularly for signals with time-varying frequency components. This prominence derives from WT's inherent capability to process nonstationary data and its rigorous mathematical foundation – specifically, its operational principle involving the convolution between the analyzed signal and adaptable filter banks [12]. This operational principle can be mathematically formalized as:
\begin{equation}W_sg(t)=g\ast\psi_s(t)=\int_{-\infty}^{\infty} g(t) \psi^*(\frac{t - b}{a}) dt,  \quad g(t) \in L^2(R).\label{eq1}\end{equation}

$g(x)$ represents input signal and g is its abbreviation; $L^2(R)$ represents square-integrable one-dimensional function Hilbert space; $\psi_s(t)$ represents the wavelet function on spaces. a represents the scale parameter, which controls the scaling of the wavelet function. b represents the translation parameter, which controls the translation of the wavelet function. It is evident from the Eq. (1) that WT essentially represents the signal $g(t)$, which is generated by a telescope filter with an impulse response of $\psi_s(t)$. Moreover, the wavelet function plays a systematic role in the process. Consequently, the spectral density corresponding to the WT in the frequency domain can be expressed as:

\begin{equation}W_sG(f)=G\cdot|\psi_s(f)|^2.\label{eq2}\end{equation}

Eqs. (1) and (2) reveal a critical constraint of WT: the mathematical duality leads to the need for a trade-off between temporal and spectral resolution governed by the uncertainty principle. This necessitates deliberate selection of wavelet bases exhibiting maximal congruence with target signal features – a process prioritizing dominant frequency capture at the expense of spectral detail. This compromise further underscores WT's intrinsic lack of adaptability, as the chosen basis must remain invariant across entire datasets despite potential nonstationarities. Thus, while WT permits basis function variation, its fundamental nature as a Fourier-analytic methodology persists [13]. Moreover, WT's reliance on subjective basis selection for orthogonal signal decomposition renders it an a priori analytical framework, where outcomes critically depend on basis judgment.

The suitability of a priori analysis in data analysis hinges largely on the depth of examination of the dataset. A meticulous and comprehensive data analysis yields more accurate a priori predictions that align closely with natural phenomena. Conversely, inadequate analysis may lead to the emergence of mathematical artifacts like disordered harmonics. Thus, whereas a priori analysis has mathematical rigor, it may lack clear physical relevance, necessitating a thorough data analysis beforehand. The emergence of EMD has largely addressed the limitations of a priori analysis, despite the controversy surrounding its lack of rigorous mathematical foundations. Nonetheless, EMD serves as an adaptive signal processing method grounded in a posteriori-principles. By utilizing extreme points and zero crossings, EMD extracts subsignals of varying amplitudes and periods within the signal, decomposing it into multiple IMFs with frequencies ranging from high to low. Unlike traditional methods, EMD bypasses integral transformations, mitigating the generation of disordered harmonics resulting from incomplete orthogonality between the chosen basis and the signal.

It has been previously mentioned that conducting a priori analysis based on specific analytical criteria is reasonable. Hence, it is natural to consider using IMFs after EMD processing as input for a prior analysis. The advantage of this approach is that the frequency range of IMFs is more homogeneous than that of the original noise signal. Despite some instances of mode mixing, most IMFs exhibit well-defined time-frequency characteristics. By using IMFs with relatively pure frequency waveform as the criterion for selection of wavelet base, better TFA results can be obtained. In summary, the specific process of EMD-WT is as follows:

\begin{figure}[htbp]
\centering\includegraphics[width=10cm]{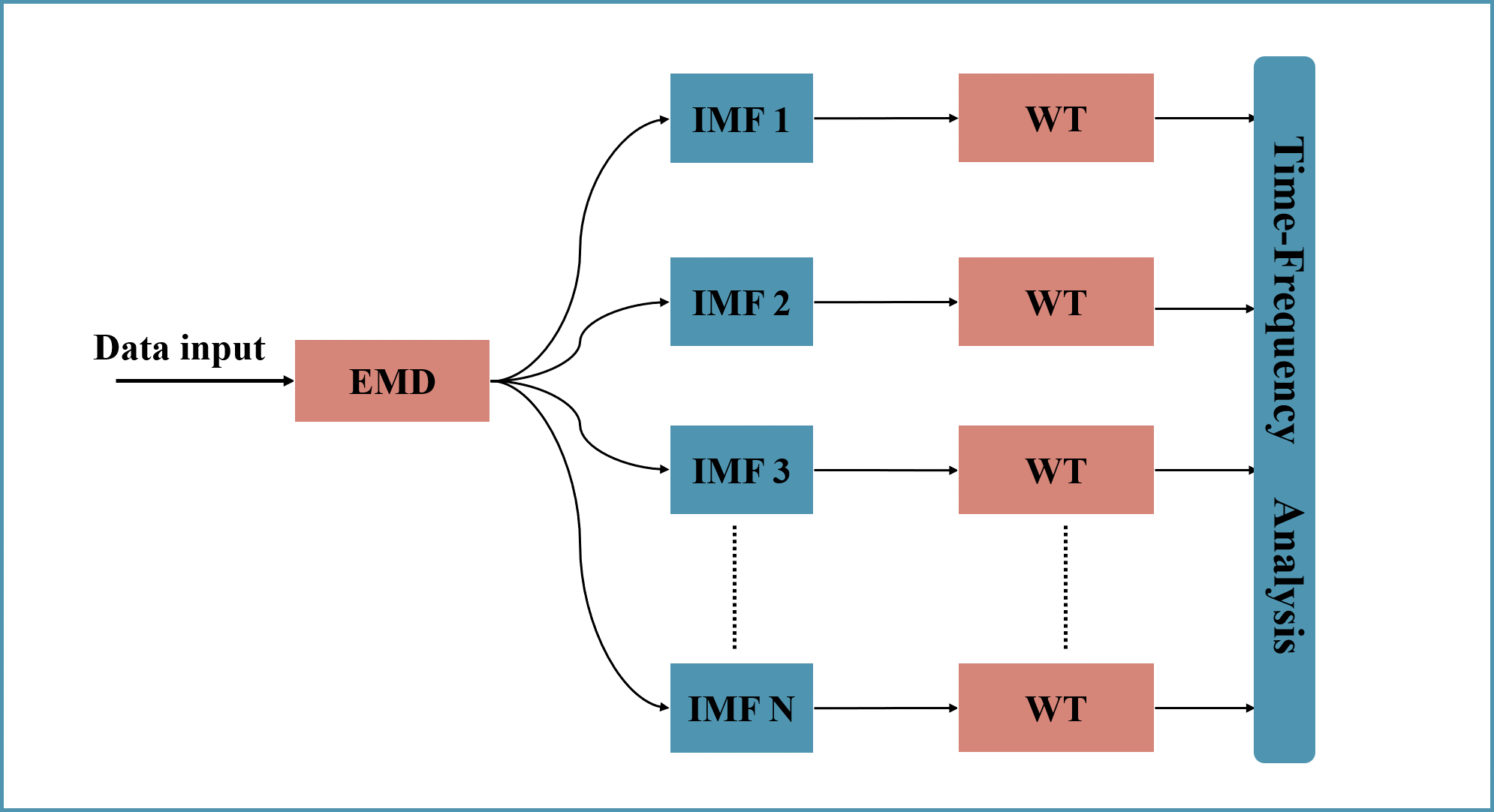}
\caption{The diagram of EMD-WT.}
\end{figure}

\subsection{Method of EMD-WT combined with ADEV}
In the realm of time-frequency transfer, the collected noise signal exhibits immense complexity, comprising various components such as frequency drift stemming from crystal aging [14], thermal noise due to temperature fluctuations, induced fast variable noise from environmental vibration (rapidly decaying damping system), locking-induced noise, and amplifier-induced noise, among others. Typically, environmental factors like temperature and vibration contribute noise closely aligned with the carrier frequency, inducing random jitter and resulting in Random-Walk Noise (RWN). The selection and design of the oscillator's physical resonant device and electronic components introduce Flicker Frequency Noise (FFN), whereas locking-induced noise generally manifests as White Frequency Noise (WFN). Amplifiers and frequency doublers predominantly introduce Flicker Phase Noise (FPN) and White Phase Noise (WPN) [15]. These noises can be seen by the slope of the ADEV, which is usually shown as -2$\sim$2. Generally, the assessment metric for frequency transfer relies heavily on ADEV, derived from the second-order structure function, limiting the number of terms in the frequency signal within the ADEV system to no more than quadratic terms. Consequently, the maximum number of noise terms in the noise modeling within this system is proportional to the square of the frequency ($f^2$ ). For transmitted frequency signals, the power law spectral noise model typically exhibits the following characteristics:

\begin{equation}S_\phi(f)=\sum_{i=-\alpha}^{\alpha}h_if^i.\label{eq3}\end{equation}

eq. (3) indicates that $S_\phi(f)\sim|f|^{-\alpha}$ and $f=\omega/2\pi$. In William C. Lindsey’s study established a subtle correlation between ADEV and the structure function [16]. It demonstrated that ADEV can be effectively expressed in terms of the structure function:

\begin{equation}\sigma^2_{\rm{ADEV}}=\frac{D^{(2)}_{\phi}(\tau)}{2\cdot(\omega_0\tau)^2},\label{eq4}\end{equation}

It shows that ADEV is essentially a second-order structure function, and that the structure function and $S_\phi(f)$ can also be transformed reciprocally, which is expressed as follows:

\begin{equation}D^{(M)}_{\phi}(\tau)=2^{2M}\int_{-\infty}^{+\infty} \sin^{2M}(\frac{\omega\tau}{2})\cdot\frac{S_\phi(\omega)}{\omega^2}\, \frac{d\omega}{2\pi}.\label{eq5}\end{equation}

When $M=2$, Eq. (5) convert to:

\begin{equation}D^{(2)}_{\phi}(\tau)=\frac{8}{\pi}\int_{-\infty}^{+\infty} \sin^4(\frac{\omega\tau}{2})\cdot\frac{S_\phi(\omega)}{\omega^2}\, d\omega.\label{eq6}\end{equation}

Considering the convergence of Eq. (6) at $\omega=0$. It converges when $\sin^4(\omega\tau/2)\sim\omega_0^4$ and $S_\phi(\omega)\sim\omega^{-\alpha}$, $\alpha\leq2$. Due to $D_\phi^{(2)}(\tau)$ is essentially ADEV, the noise power spectral density of ADEV does not exceed second order.

In summary, the type of noise can be inferred from the slope of the ADEV, however, the entire dataset can be masked by one or two dominant noises, so analysis based solely on ADEV is not comprehensive. Multiple IMFs decomposed by EMD can be used for ADEV analysis, so that noise type analysis can be performed according to different frequency bands. Further, by combining the analysis results with the EMD-WT method, this approach allows us to discern not only the type of noise but also its time-frequency characteristics.

\subsection{Method of wavelet selection}

It is essential to select wavelet bases in WT, and the selection of wavelet basis is conducive to better time-frequency analysis. The selection of a wavelet basis typically involves a wavelet with moderate support length, symmetry, and regularity, a high vanishing moment, and similarity to the signal waveform [17, 18, 19]. Based on experimental experience and the characteristics of wavelet bases, the Haar, Meyer, and Morlet wavelet bases are selected for comparison. These three wavelet bases are chosen for their regularity, symmetry, and similarity to signal waveforms. Fig.2 shows the signal representation of IMF-1 used for comparison across three wavelet bases.

\begin{figure}[htbp]
\centering\includegraphics[width=9cm]{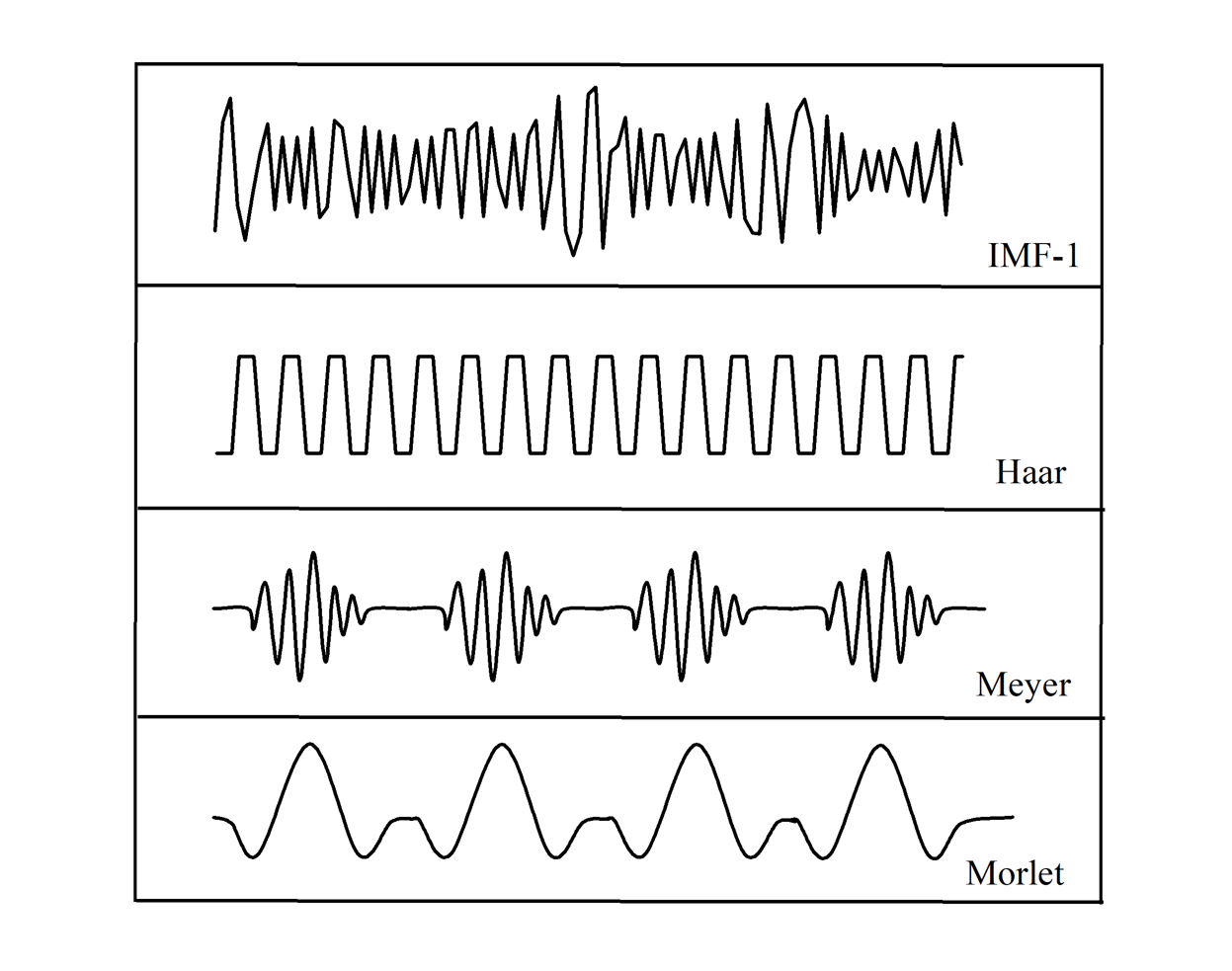}
\caption{Wavelet Basis Comparison for IMF-1.}
\end{figure}

As demonstrated in Fig. 2, the Meyer and Morlet wavelet bases exhibit significantly closer alignment with the original oscillatory characteristics of IMF-1 (derived from EMD decomposition) compared to the Haar wavelet basis. This empirical observation justifies the preferential adoption of Meyer or Morlet wavelet bases for practical data analysis applications. Extending this methodology systematically, each IMF can undergo individualized wavelet basis selection through an analogous optimization process. This adaptive strategy effectively circumvents the previously discussed limitations of conventional WT in handling multimodal signals, specifically its inherent lack of flexibility in basis adaptation. The proposed workflow, schematically illustrated in the accompanying figure, operates as follows:

\begin{figure}[htbp]
\centering\includegraphics[width=9cm]{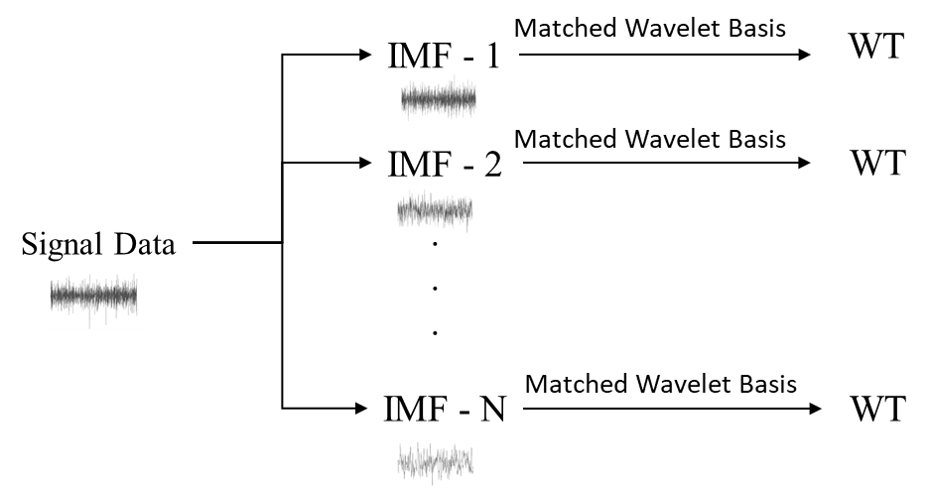}
\caption{Flowchart of EMD-WT method.}
\end{figure}

\section{Experiment}
\subsection{The experimental setup of phase noise data collection of mode locking and EDFA}
The first step of this study involves the use of locking and EDFA noise data to demonstrate the advantages of the ADEV combined with the EMD-WT method. This is due to the following reasons: 1. The locking data is suitable for comparing the EMD-WT, WT, and HHT methods, as the noise in the locking data is relatively simple, making it easier to decompose using all three methods. 2. Since the locking data contains only a single type of noise, it does not fully showcase the advantages of the ADEV combined with the EMD-WT method. Therefore, EDFA data is required for demonstration and illustration (To avoid confusion, it should be clarified that the EDFA noise data refers to data where EDFA noise is superimposed on locking noise.). The data collection is divided into two parts:

a. The synchronization noise, specifically mode-locking noise, is analyzed independently. This analysis involves employing the EMD-WT method after phase acquisition, which is achieved by synchronizing the atomic clock signal with the mode-locking signal.

b. The mode-locking signal is split into two paths: one transmitting only the mode-locking signal, and the other transmitting the signal after it passes through the device. The phase mixing of these signals is collected, and EMD-WT is utilized to analyze the noise introduced by the combined effect of mode-locking and the device.

The noise collection mentioned above was primarily conducted using a mixing technique. The original signal comes from the clock path and is transmitted directly, while the synchronized laser is transmitted on a separate path. After transmission, the laser is mixed with the clock signal and passed through a low-pass filter, with the filtered data corresponding to mode-locking noise. For the EDFA noise, after the laser passes through the EDFA, it is mixed with the clock signal and filtered. The resulting signal contains a mixture of mode-locking noise and EDFA noise.

\begin{figure}[htbp]
\centering\includegraphics[width=10cm]{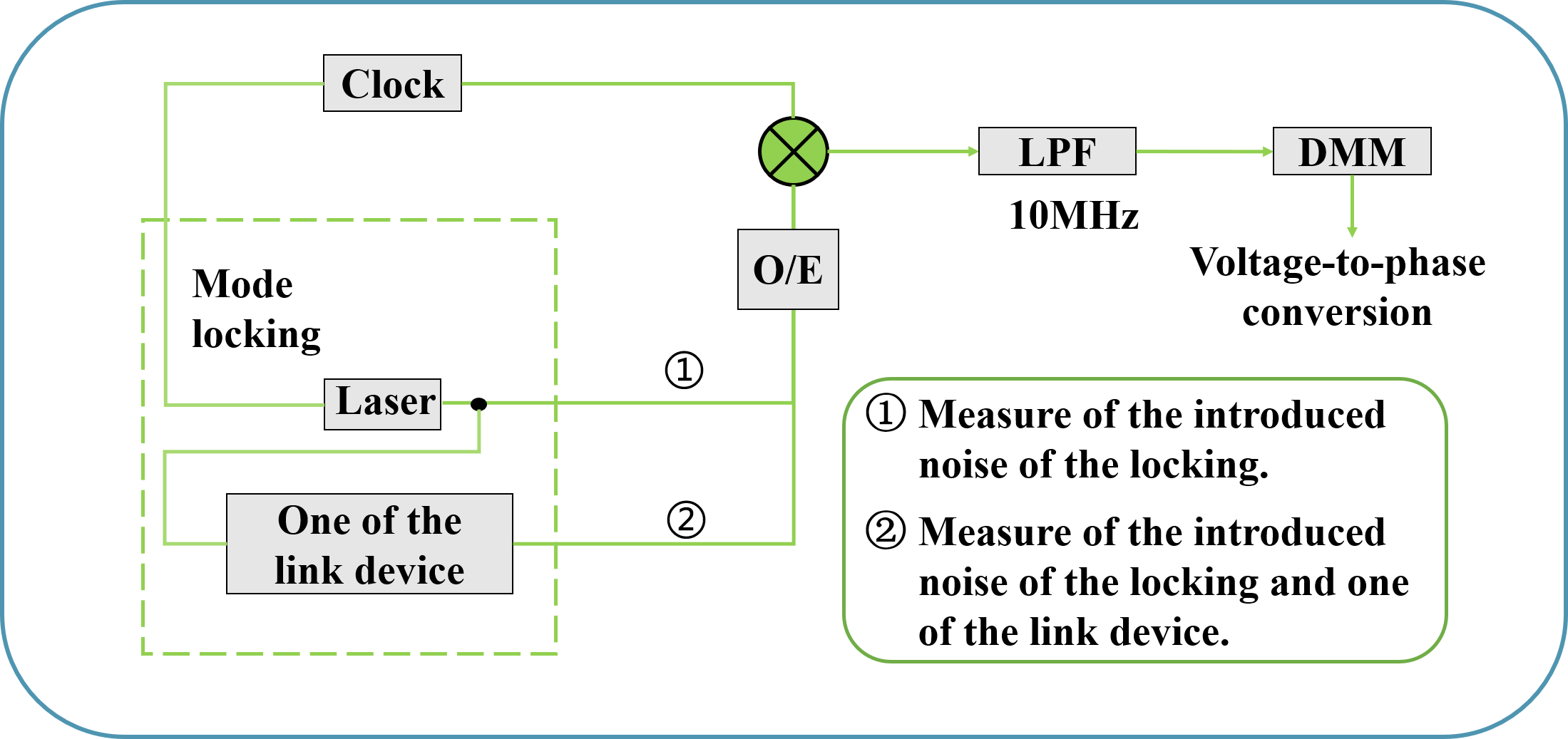}
\caption{The local analysis structure of the link(DMM represents digital multi meter; LPF represents low-pass filter; O/E represents optical-electrical conversion system).}
\end{figure}

Finally, the time-frequency graphs obtained from both parts (a and b) are compared with ADEV to discern the noise types and characteristics of the device. The experimental setup for this analysis is depicted in Fig. 4. 

In this paper, the device in Fig. 4 uses Erbium-doped Optical Fiber Amplifier (EDFA), and the sampling frequency is 1Hz. The noise analysis of EDFA in frequency transfer system is the most complex compared with other devices, so if the noise characteristics of EDFA can be clearly analyzed, EMD-WT can be also applied to other devices in the system. 

For the phase deviation acquisition, the voltage data are measured using a DMM after the signal is mixed and passed through a LPF. The phase deviation is then obtained through a voltage-to-phase conversion.

\subsection{The experimental setup of phase noise data collection of frequency transfer link}
The preceding locking and EDFA noise datasets are primarily intended to demonstrate the advantages of the ADEV combined with EMD-WT analysis. In addition, the method needs to be applied to the entire time-frequency transfer link for analysis. A complete time-frequency transfer link is illustrated in the figure below:

\begin{figure}[htbp]
\centering\includegraphics[width=10cm]{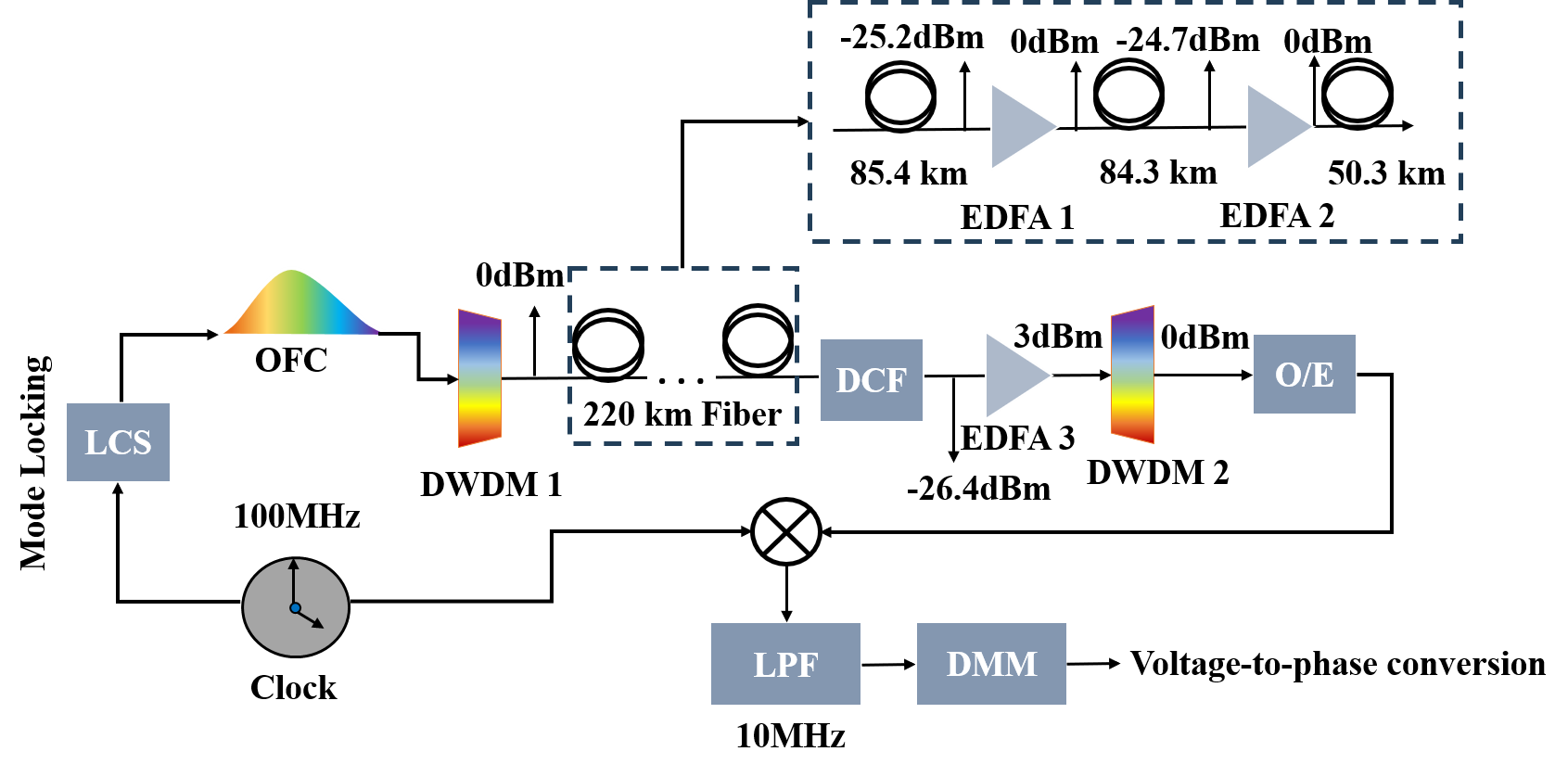}
\caption{The experimental structure of frequency transfer link (LCS represents locking control system; OFC represents optical frequency comb; DCF represents dispersion compensation fiber; DWDM represents dense-wavelength division multiplexer).}
\end{figure}

Fig. 5 illustrates that the OFC locks the clock frequency through the LCS. Due to the multiple wavelengths of the OFC, DWDM is used for wavelength selection, with the C32 channel chosen for this experiment. The repetition frequency of the OFC is locked to a 100 MHz reference signal. The light then passes through a 220 km optical fiber (standard single-mode fiber), and the chromatic dispersion-induced frequency deviation is compensated by the commercial DCF modules (dispersion parameter:-100 ~-250 ps/nm·km, about 15 \textasciitilde 37.4km). Following amplification by the amplifier and O/E conversion (the O/E module used is composed of a DC-type photodiode (PD) with a bandwidth of 100 MHz), the signal is mixed with the clock, and phase deviation data is collected, which corresponds to the link noise data. 

In addition, since the experiment requires that the input optical power of each EDFA be controlled between –30 dBm and –20 dBm, we initially set the optical power to 0 dBm. After 85.4 km of fiber transmission, the input power to EDFA 1 is –25.2 dBm and is amplified back to 0 dBm. After a further 84.3 km of transmission, the input power becomes –24.7 dBm and is again amplified to 0 dBm by EDFA 2. It then propagates through commercial DCF modules(about 15 \textasciitilde 37.4km), with a loss of about 13 dBm. After an additional 50.3 km of fiber, the input power to EDFA 3 is –26.4 dBm. However, since DWDM 2 is required for filtering to reduce noise and improve optical isolation, this process introduces additional power loss. Therefore, EDFA 3 needs to amplify the signal to 3 dBm, so that the input power to the O/E module can be maintained at 0 dBm.

\subsection{The comparative experimental setup of WT, HHT and EMD-WT}

To rigorously validate the comparative efficacy of WT, EMD, and the hybrid EMD-WT method – particularly when integrated with the ADEV metric – a systematic experimental architecture has been designed. As schematically illustrated in Fig. 6, the experimental workflow proceeds as follows:

\begin{figure}[htbp]
\centering\includegraphics[width=9cm]{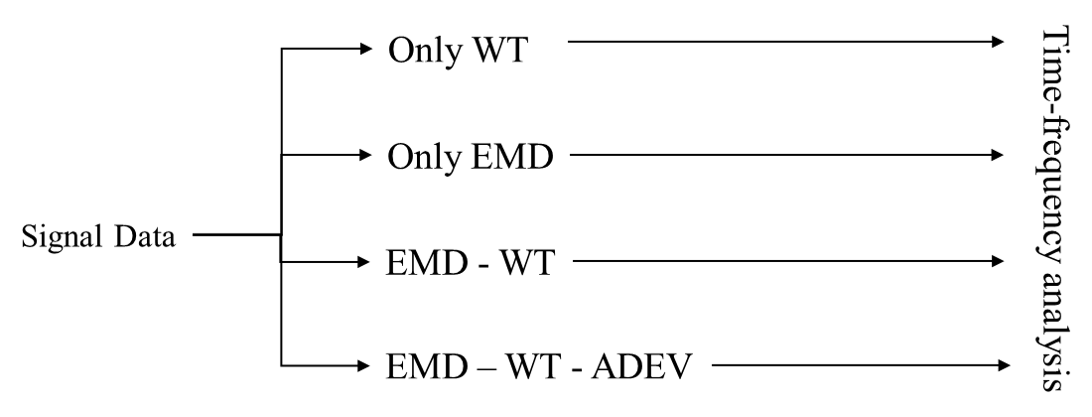}
\caption{The comparative experimental structure of the four methods.}
\end{figure}

The following points need to be noted: 1. Input Signal Standardization: Identical raw signal datasets, representative of frequency transfer system outputs under varying noise conditions, serve as unified inputs to all analytical methodologies. 2. Algorithmic Parallel Processing: baseline WT implementation: Conventional wavelet decomposition using predetermined basis functions. Standalone EMD analysis: Adaptive decomposition without wavelet integration. Hybrid EMD-WT Framework: EMD-derived IMFs subjected to component-specific wavelet basis selection. ADEV-Enhanced EMD-WT: Integrated analysis combining time-frequency representations with ADEV noise characterization

\section{Result}

\subsection{The three methods – WT, HHT and EMD-WT}

First, the locking noise data is used to compare the three methods, as the locking noise is simple and facilitates the decomposition by all three methods, resulting in more intuitive outcomes. For WT, if the data is not stratified, the larger noise component will cover up other components with lower noise intensity, resulting in inaccurately noise analysis. To analyze the noise of each frequency band requires accurate adjustment of two parameters of WT, which lacks adaptability. Taking the locking noise as an example (structure \normalsize{\textcircled{\scriptsize{1}}} in Fig. 4), the result of WT for the whole locking noise is given:

\begin{figure}[htbp]
\centering\includegraphics[width=10cm]{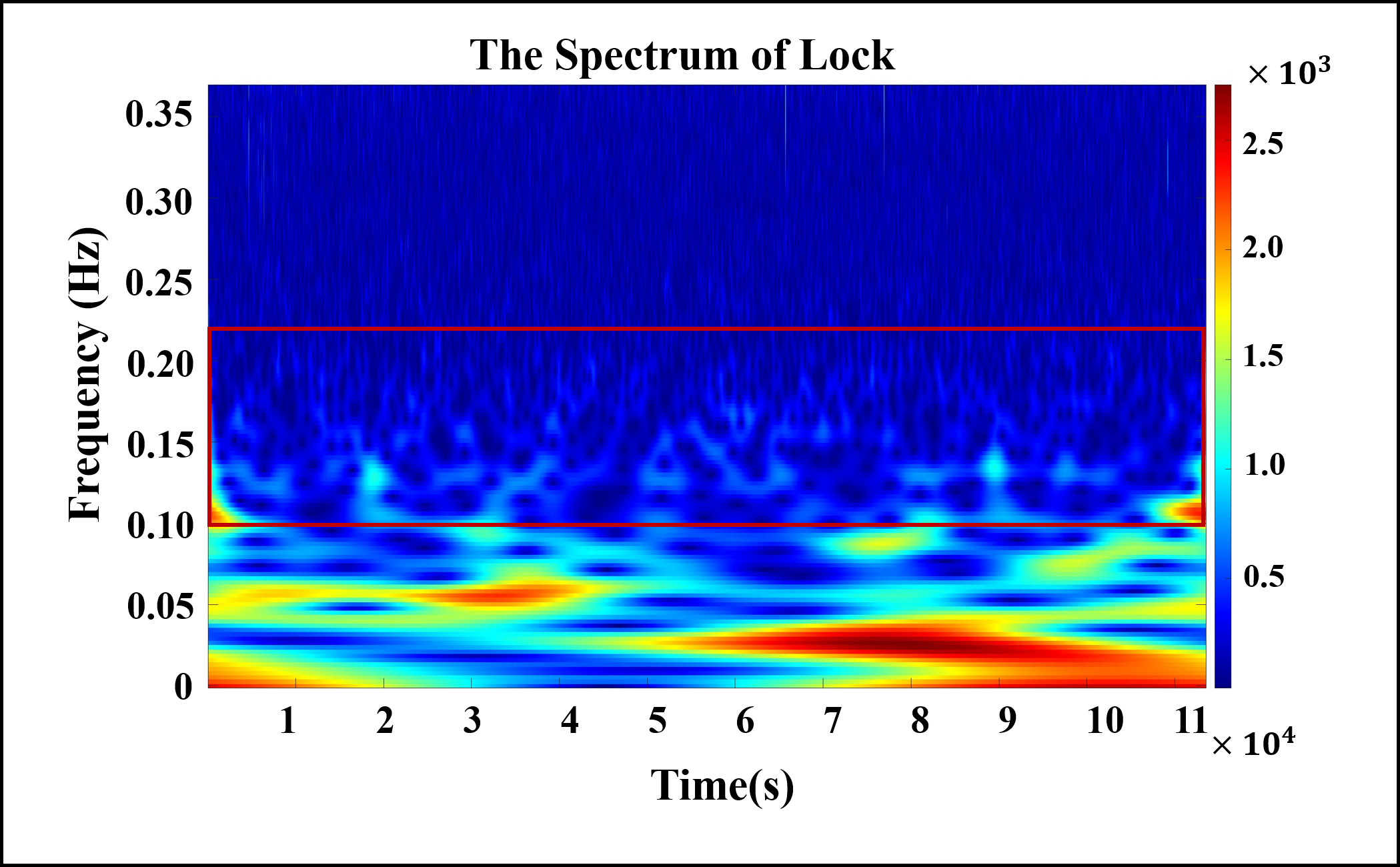}
\caption{WT of locking noise data.}
\end{figure}

As can be seen in Fig. 7, there is a noise component in the red box. However, because the noise component in the range of [0, 0.05Hz] is too strong, some noise components in other bands are covered up, which leads to inaccuracy of the analysis results. To obtain accurate analysis of the frequency band in the red box, the parameters of the wavelet basis need to be adjusted, so the adaptability is lost.

Due to the high precision of the frequency transfer signal, even a small mechanical vibration can cause high-frequency jitter, resulting in signal discontinuity. In the HHT, mode mixing in the initial IMFs disperses noise across the time-frequency graph, leading to analysis inaccuracies. To address this problem, four IMFs are selected from the locked data for Hilbert transform, constituting one HHT.

\begin{figure}[htbp]
\centering\includegraphics[width=10cm]{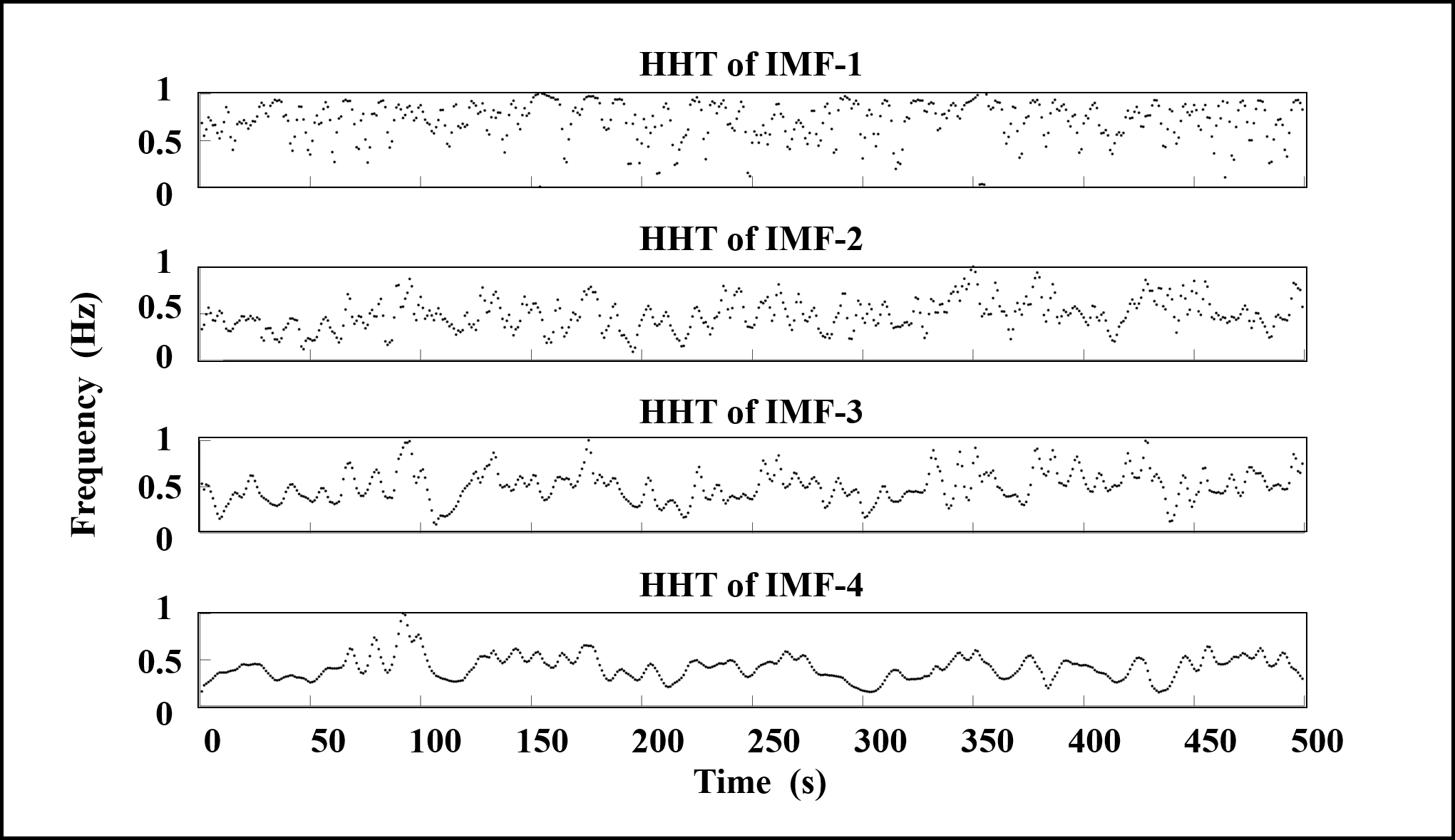}
\caption{Hilbert spectrum of IMF1$\sim$4.}
\end{figure}

The reason for selecting these four layers of IMF is to visualize the mode mixing phenomenon caused by using only the EMD method. In Fig. 8, the HHT of IMF1 and IMF2 clearly shows mode mixing of each noise component, with the noise dispersed across different frequency bands. IMF3 and IMF4, on the other hand, exhibit slight mode mixing. This indicates that the use of the EMD method alone leads to mode mixing in the first few layers of IMF.

Hence, to address problems arising from WT and HHT, EMD-WT can be employed for noise analysis. Subsequently, the following section presents EMD-WT results of the locking data.

\begin{figure}[htbp]
\centering\includegraphics[width=10cm]{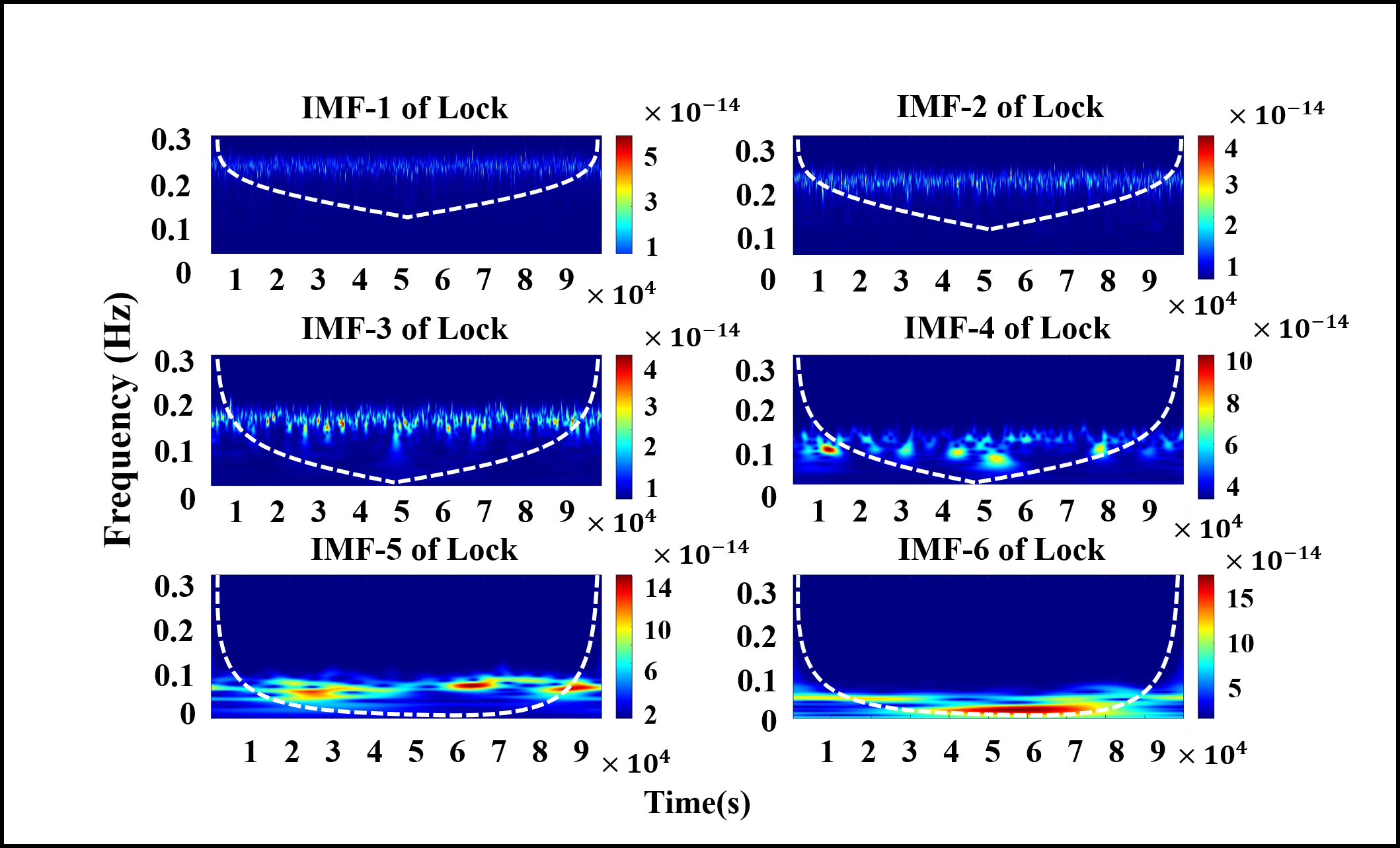}
\caption{IMFs of EMD-WT (The noise level of lock is $10^{-14}$).}
\end{figure}

The noise components around 0.3Hz, 0.25Hz, 0.2Hz, 0.15Hz, 0.10Hz and 0.05Hz are selected for the six IMFs respectively (The white dashed line in Fig. 9 represents the Cone of Influence (COI). It can be observed that most of the energy in the frequency bands of the six locked IMFs lies within the COI, indicating that the majority of the obtained results are reliable). 

The main reason for choosing 6-layer IMFs is that the other layers of IMFs contain the same frequency. Therefore, in order to simplify the image representation, the IMFs are extracted into the main 6 frequency bands. It can be seen that the frequency is clearly arranged in the range from high to low, which effectively solves the shortcomings of WT and HHT in analyzing high-precision and mixed multiple noise signals. Comparing Figs. 7 and 9, it is not difficult to find that EMD-WT has the advantage of displaying details and more flexible in different frequency bands compared with WT.  Comparing Figs. 8 and 9, it is not difficult to find that EMD-WT has the advantage of avoiding mode mixing compared with HHT.

\subsection{Noise type and characteristic analysis -- EMD-WT combined with ADEV analysis}
Based on the aforementioned analysis, ADEV can discern the noise type through its slope, but ADEV analysis of whole data will only show one or two slopes, which cannot effectively analyze the noise introduced by devices of the system (Fig. 10 shows only tends -1 slope). whereas EMD-WT can identify the frequency band and corresponding time of the noise. Therefore, these two methods can be effectively integrated to analyze not only the predominant noise type but also its time-frequency characteristics. The EDFA noise data is used for the ADEV combined with EMD-WT analysis.  Since the EDFA noise data contains multiple types of noise, the EMD-WT method can effectively separate the noise at different frequencies and use ADEV to obtain information about the types of noise.

\begin{figure}[htbp]
\centering\includegraphics[width=10cm]{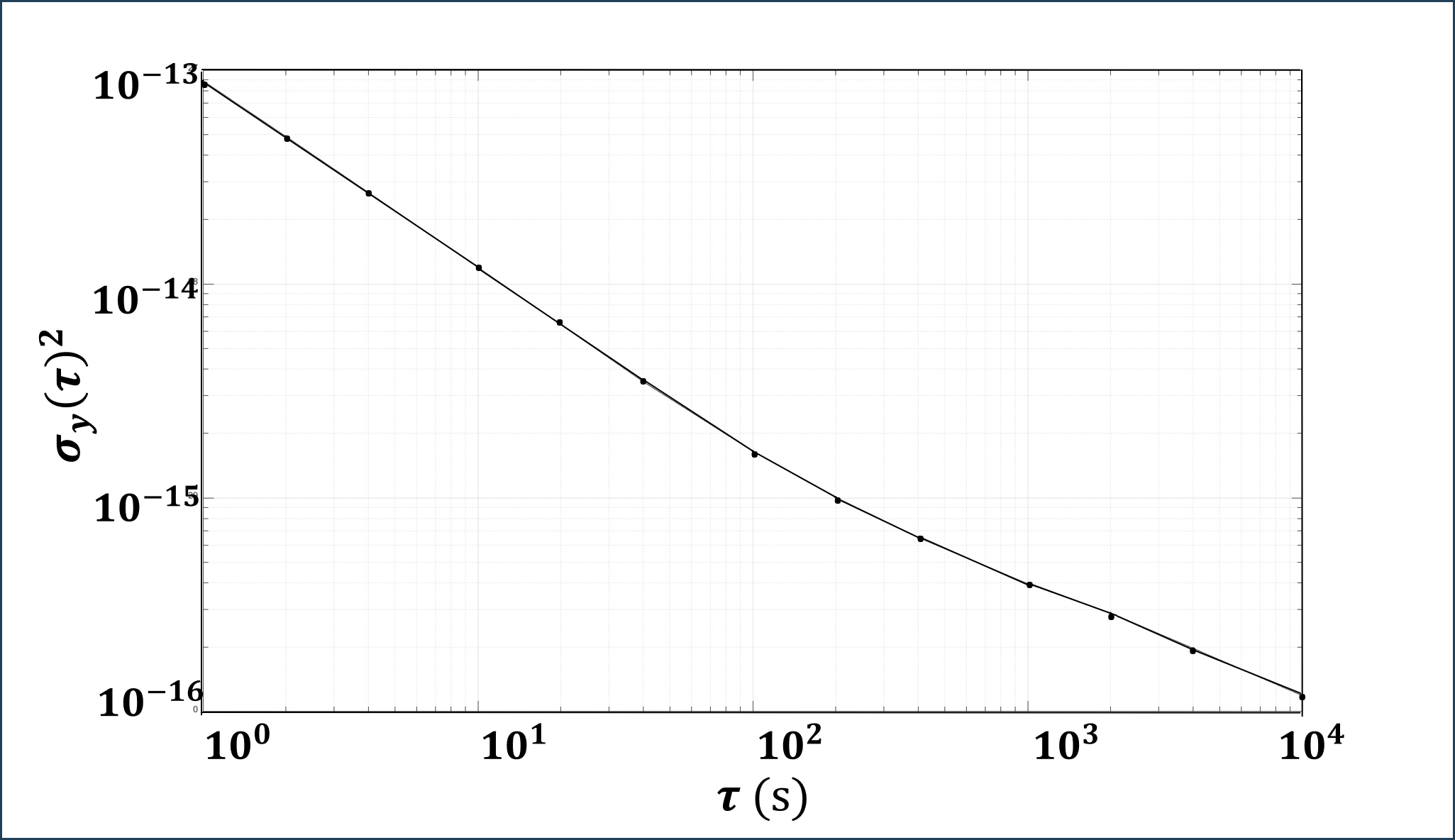}
\caption{ADEV of the whole data.}
\end{figure}

For instance, considering the EDFA within the time-frequency transfer system. Structure \normalsize{\textcircled{\scriptsize{2}}} in Fig. 4 is replaced with EDFA. Figs. 11 and 12 depict the time-frequency diagram and ADEV of IMFs (IMF5$\sim$8) at the median frequency of the EDFA.It can be seen that the frequency of IMF 5$\sim$8 is in the range of [0.15Hz, 0.25Hz] (As can be seen in Fig. 12,  the frequency components of the four IMFs corresponding to the EDFA are largely within the COI. Therefore, the wavelet analysis results obtained from these components are considered to be highly reliable), and the corresponding ADEV slope tends to -2 over a long period of time, which indicates that the dominant components of noise are WPN, FPN. That is, the noise dominant component is introduced by EDFA.

\begin{figure}[htbp]
\centering\includegraphics[width=10cm]{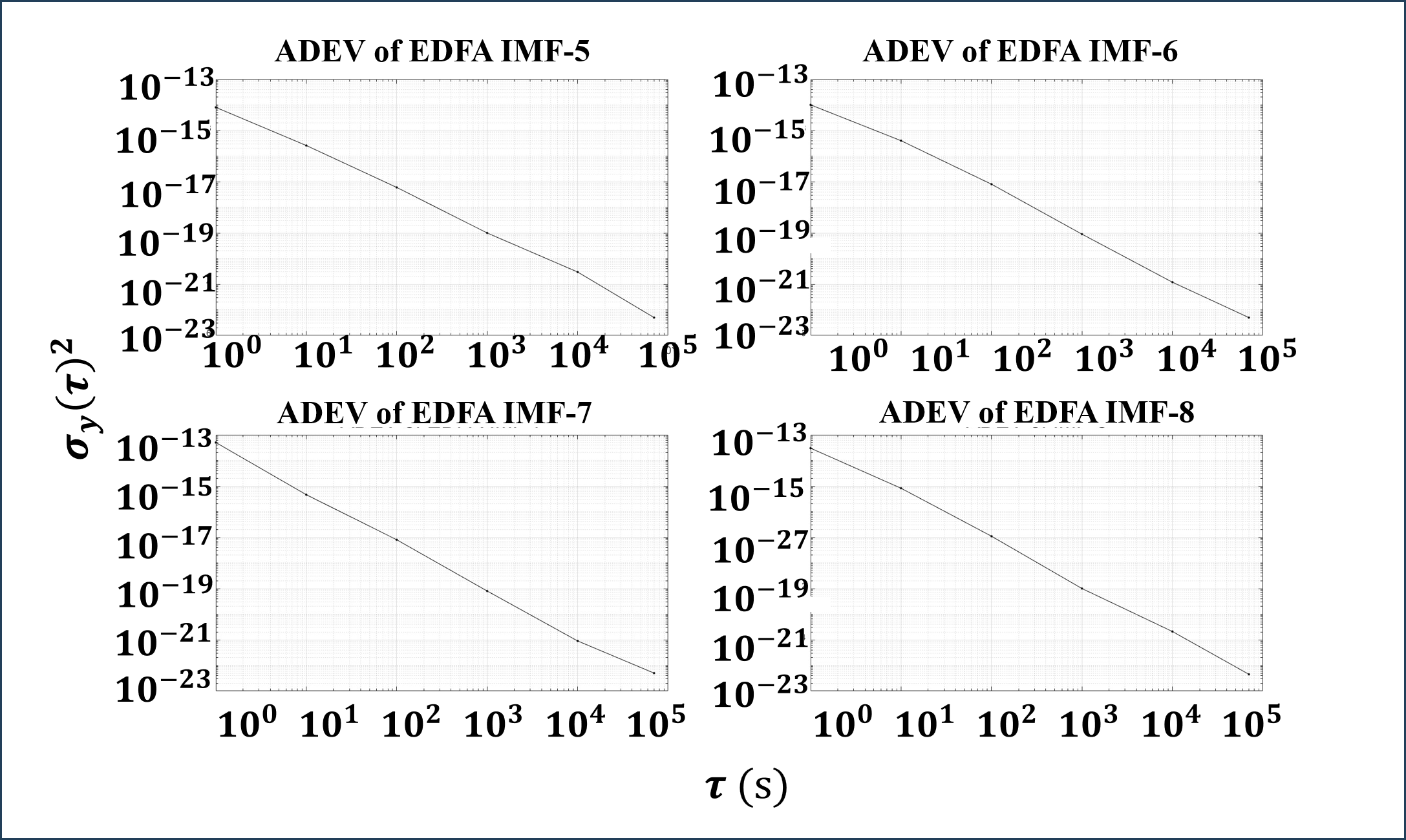}
\caption{ADEV of EDFA (IMF5$\sim$8).}
\end{figure}

\begin{figure}[htbp]
\centering\includegraphics[width=10cm]{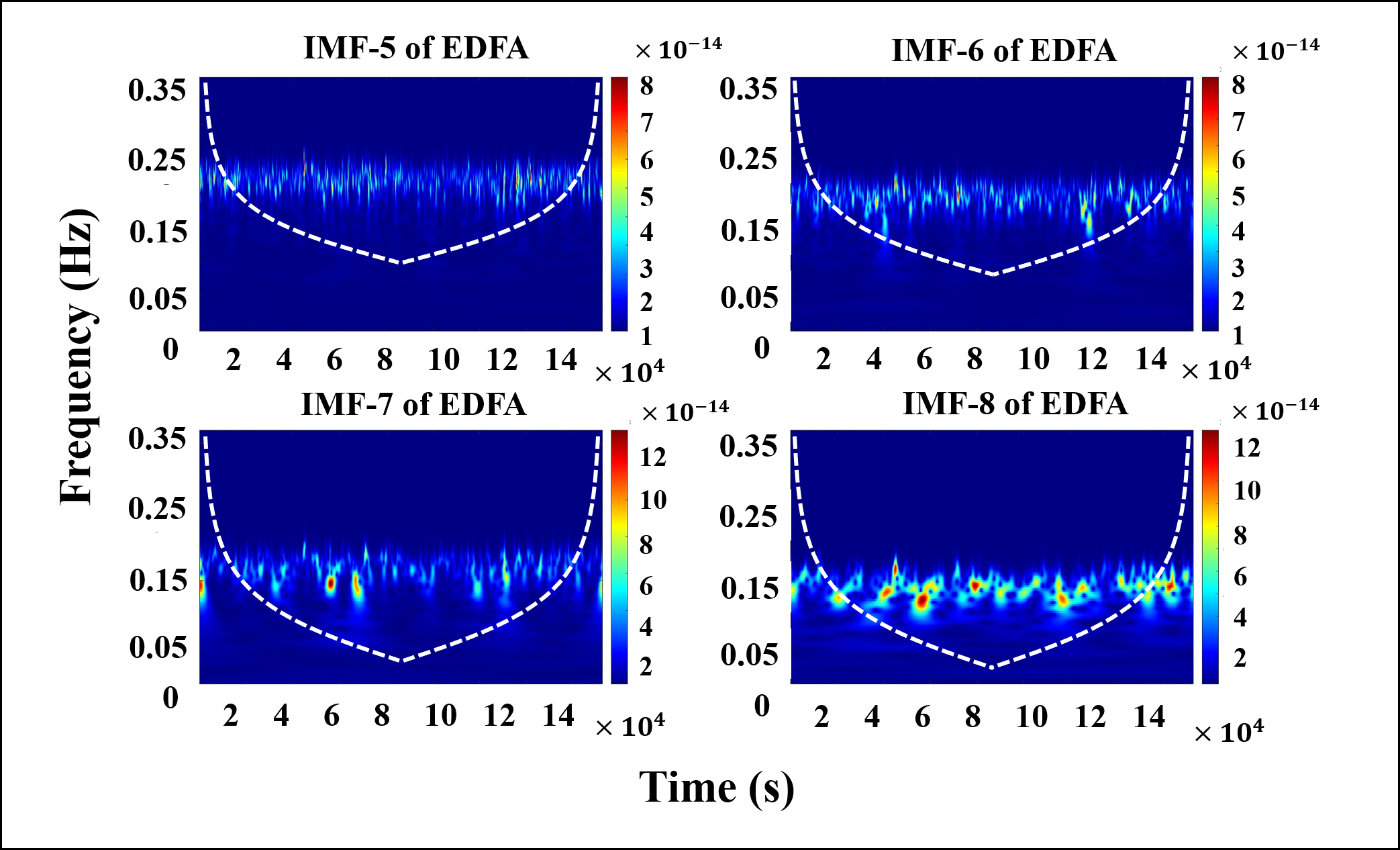}
\caption{Time-frequency graph of EDFA (IMF5$\sim$8).}
\end{figure}

It can be observed that the ADEV combined with EMD-WT analysis effectively analyzes the types of noise introduced by the EDFA and their corresponding frequency bands.  Similarly, if there is a need to independently analyze other devices in the frequency transfer system, this method can also be used for a reasonable analysis.

\subsection{The analysis of frequency transfer system}
The advantages of combining EMD and WT with ADEV have been analyzed above. The noise data of the entire link is then processed, and the corresponding EMD+WT analysis results are shown in Fig. 13:

\begin{figure}[htbp]
\centering\includegraphics[width=10cm]{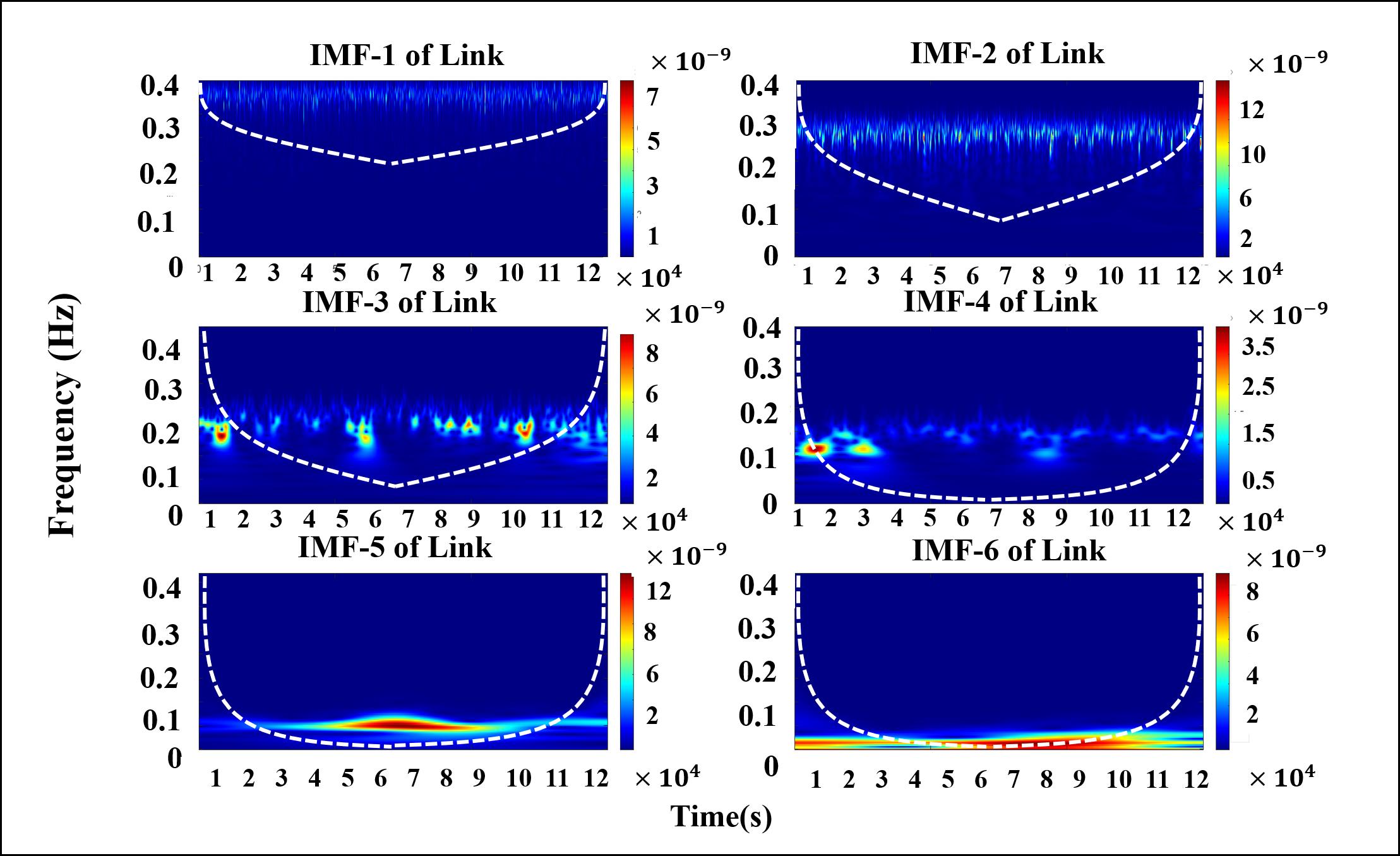}
\caption{IMFs of frequency transfer Link.}
\end{figure}

After decomposition, the entire link yields 14 IMFs. However, several of these layers contain overlapping frequency components. Therefore, one IMF is selected from each duplicated frequency band for presentation. In total, six IMFs are chosen, it can be observed that the frequency components of IMF1–5 for the link are mostly within the COI. Although only a small portion of the energy in IMF6 lies within the COI, this portion has relatively low energy, while the majority of the relevant components remain inside the COI. Therefore, the resulting analysis can be considered reliable. Their corresponding ADEV plots are shown in Fig. 14.

\begin{figure}[htbp]
\centering\includegraphics[width=10cm]{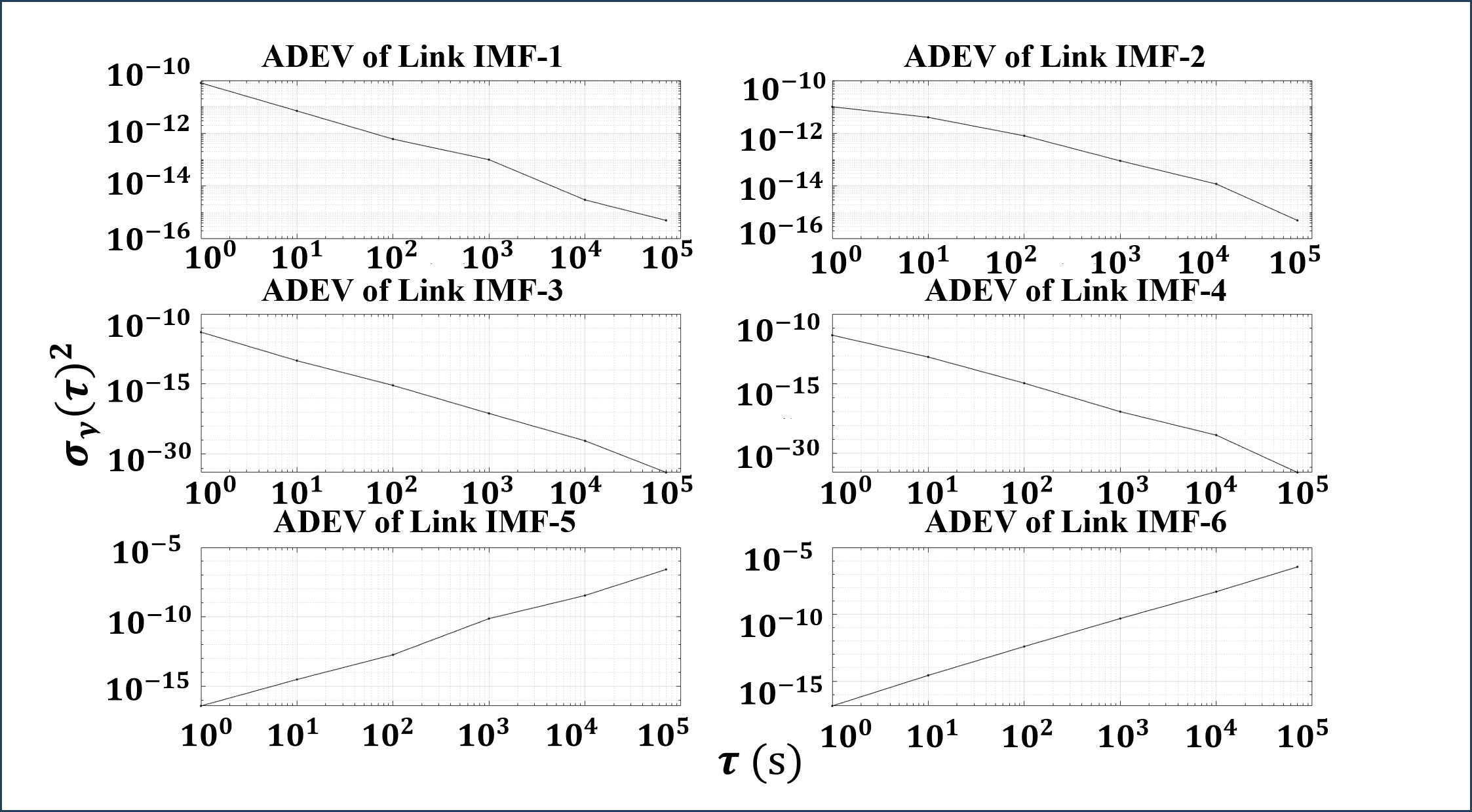}
\caption{ADEVs of frequency transfer Link.}
\end{figure}

It can be observed that in the higher frequency bands (IMF1 and IMF2), the ADEV exhibits a slope of –1, indicating that the dominant noise type in these bands is WFN. In the mid-frequency bands (IMF3 and IMF4), the ADEV shows a slope of –2, suggesting that the noise is primarily governed by FPN and FWN, which are introduced by the EDFA in the link. Finally, in the lower frequency bands (IMF5 and IMF6), the ADEV displays a slope of +2, which corresponds to frequency drift. This phenomenon is mainly attributed to the aging rate of the atomic clock itself.

The experimental results conclusively demonstrate the analytical superiority of the EMD-WT-ADEV framework over conventional WT, standalone EMD, and the EMD-WT hybrid approach in characterizing time-frequency transfer systems. As clearly evidenced by the comparative data presented in Fig. 15, the integrated methodology achieves enhanced performance across three critical dimensions:

\begin{figure}[htbp]
\centering\includegraphics[width=9cm]{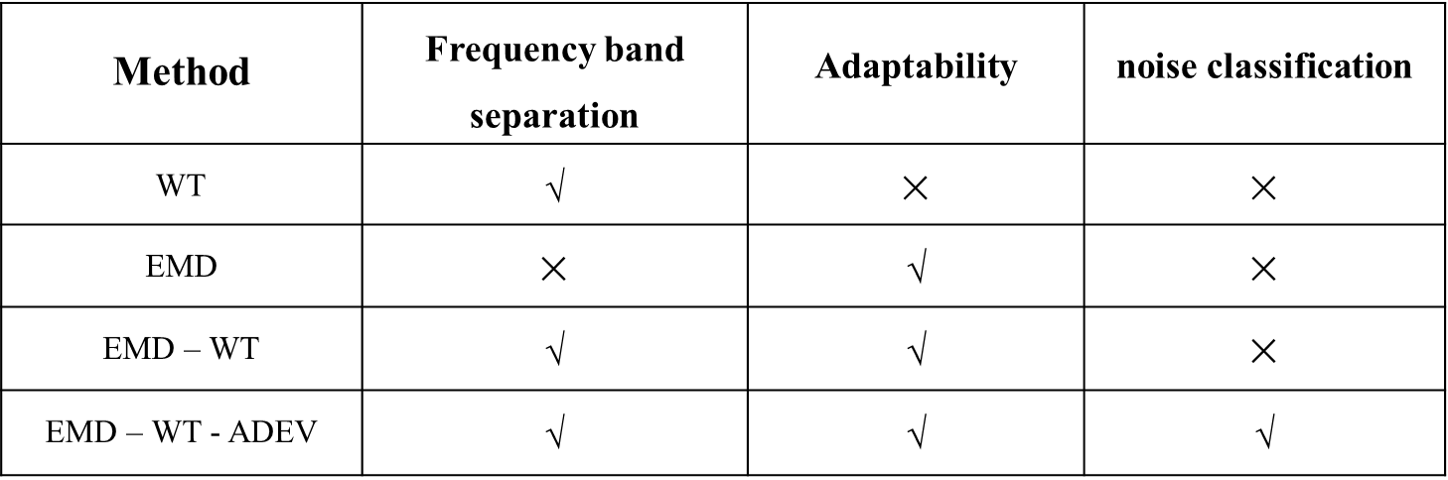}
\caption{Ability of four methods in frequency transfer systems.}
\end{figure}

\section{Conclusion}

The innovation of this study lies in the first-time introduction of EMD-WT into time-frequency transfer systems, combined with ADEV to perform a joint noise analysis of the two core components in the frequency transfer system — EDFA and mode-locking noise. By combining these with the ADEV, a conventional evaluation metric in the field, a comprehensive framework for system noise characterization is established. Initially, we analyze the inherent limitations of standalone EMD and WT, demonstrating the rationale for their combined use in time-frequency feature analysis. The application of EMD as a preprocessing step enables component-specific wavelet basis selection for each IMF, thereby fulfilling the adaptability requirements of time-frequency analysis. The integration of EMD-WT with ADEV significantly enhances the accuracy of system evaluation.

The EMD-WT-ADEV methodology fundamentally addresses two critical challenges: 1. Incompleteness of evaluation metrics in time-frequency transfer systems. 2. Lack of prior knowledge due to the inability of conventional TFA methods to distinguish noise types.

By leveraging complementary advantages, this approach overcomes these limitations: ADEV analyzes noise composition through slope variations, while EMD-WT identifies noise periodicity and frequency information, collectively enabling complete noise characterization.

Finally, noise data from the frequency transfer system was selected for experimental verification. The results confirmed the feasibility of the EMD-WT-ADEV framework, demonstrating that this analytical method can not only be applied to the analysis of individual components within a system but also to the analysis of the entire frequency transfer system.

\section{Back matter}

\begin{backmatter}

\bmsection{Disclosures}
The authors declare no conflicts of interest.

\bmsection{Data availability} 
Data underlying the results presented in this paper are not publicly available at this time but may be obtained from the authors upon reasonable request.

\end{backmatter}




\end{document}